\documentclass[twocolumn]{aastex61}

\newcommand{\quasar}{P352--15}

\newcommand{\hi}{H\ensuremath{\,\textsc{i}}}

\newcommand{\nv}{N\ensuremath{\,\textsc{v}}}

\newcommand{\ips}{\ensuremath{i_{\rm P1}}}

\newcommand{\zps}{\ensuremath{z_{\rm P1}}}

\newcommand{\yps}{\ensuremath{y_{\rm P1}}}

\accepted{May 8, 2018}

\shorttitle{Powerful radio-loud quasar at $z\sim 6$}
\shortauthors{Ba\~nados et al.}

\begin{document}

\title{A powerful radio-loud quasar at the end of cosmic reionization}

\correspondingauthor{Eduardo Ba\~nados}
\email{ebanados@carnegiescience.edu}

\author[0000-0002-2931-7824]{Eduardo Ba\~nados}
\altaffiliation{Carnegie-Princeton Fellow}
\affiliation{The Observatories of the Carnegie Institution for Science, 813 Santa Barbara Street, Pasadena, CA 91101, USA}

\author[0000-0001-6647-3861]{Chris Carilli}
\affiliation{National Radio Astronomy Observatory, Pete V. Domenici Array Science Center, P.O. Box 0, Socorro, NM 87801, USA}
\affiliation{Astrophysics Group, Cavendish Laboratory, JJ Thomson Avenue, Cambridge CB3 0HE, UK}

\author[0000-0003-4793-7880]{Fabian Walter}
\affiliation{{Max Planck Institut f\"ur Astronomie, K\"onigstuhl 17, D-69117, Heidelberg, Germany}}
\affiliation{National Radio Astronomy Observatory, Pete V. Domenici Array Science Center, P.O. Box 0, Socorro, NM 87801, USA}

\author[0000-0003-3168-5922]{Emmanuel Momjian}
\affiliation{National Radio Astronomy Observatory, Pete V. Domenici Array Science Center, P.O. Box 0, Socorro, NM 87801, USA}

\author[0000-0002-2662-8803]{Roberto Decarli}
\affiliation{INAF -- Osservatorio di Astrofisica e Scienza dello Spazio, via Gobetti 93/3, I-40129, Bologna, Italy}

\author[0000-0002-6822-2254]{Emanuele~P.~Farina}
\affiliation{Department of Physics, Broida Hall, University of California, Santa Barbara, CA 93106--9530, USA}

\author[0000-0002-5941-5214]{Chiara Mazzucchelli}
\affiliation{{Max Planck Institut f\"ur Astronomie, K\"onigstuhl 17, D-69117, Heidelberg, Germany}}

\author[0000-0001-9024-8322]{Bram P. Venemans}
\affiliation{{Max Planck Institut f\"ur Astronomie, K\"onigstuhl 17, D-69117, Heidelberg, Germany}}

\begin{abstract}
We present the discovery of the radio-loud quasar PSO~J352.4034--15.3373 at $z=5.84\pm 0.02$. This quasar is the radio brightest source known, by an order of magnitude, at $z\sim 6$ with a flux density in the range of $8-100$\,mJy from 3\,GHz to 230\,MHz and a radio loudness parameter $R=f_{\nu,5{\rm GHz}}/f_{\nu,4400{\rm A}}\gtrsim 1000$. This source provides an unprecedented opportunity to study powerful jets and radio-mode feedback at the highest redshifts, and presents the first real chance to probe deep into the neutral intergalactic medium by detecting 21 cm absorption at the end of cosmic reionization.
\end{abstract}

\keywords{cosmology: observations --- cosmology: early universe  --- quasars: individual (PSO~J352.4034--15.3373)}



\section{Introduction} \label{sec:intro}

Powerful radio sources are important laboratories for the study of the most massive galaxies \citep{seymour2007}, the densest cosmic environments \citep{venemans2007b}, and test cosmological models (e.g., \citealt{caoS2017}). Furthermore, powerful radio-jets are thought to play a key role for the formation and growth of supermassive black holes, allowing super-Eddington accretion rates (e.g., \citealt{volonteri2015}). Importantly, the identification of bright radio sources at the highest redshifts promises to provide fundamental information about the epoch of reionization by detecting 21 cm absorption due to the intervening neutral intergalactic medium (e.g., \citealt{carilli2002,carilli2007b,semelin2016}).  A major limitation for such studies is that it is not clear whether those bright radio sources in the epoch of reionization exist at all \citep{saxena2017,bolgar2018}.
Although the radio-loud fraction of quasars does not seem to evolve up to $z\sim 6$ \citep{banados2015a}, thus far there is a dearth of the most extreme radio sources at the highest redshifts.

In this Letter we present the discovery of a $z\sim 6$ quasar with a radio emission that is one order of magnitude brighter than any other radio source known so far at those redshifts. We discuss the optical and radio properties of this powerful radio-loud quasar, its implications, and possible future applications. In a companion paper we present resolved VLBA observations of the quasar radio emission and discuss its jet properties \citep{momjian2018}.   We use a flat cosmology
with $H_0 = 70 \,\mbox{km\,s}^{-1}$\,Mpc$^{-1}$, $\Omega_M = 0.3$, and $\Omega_\Lambda = 0.7$. Photometric magnitudes are reported in the AB system.

\section{A radio-loud quasar at $\lowercase{z}\sim 6$}

\subsection{Selection and discovery}

PSO~J352.4034--15.3373 (hereafter \quasar) was selected as a $z\sim$ 6 quasar candidate from the Panoramic Survey Telescope \& Rapid Response System 1 \citep[Pan-STARRS1;][]{chambers2017} survey as part of the search presented in
\cite{banados2014, banados2016}. In short, we required a drop-in flux between the \ips\ and \zps\ bands ($\ips - \zps > 2$) and a flat continuum ($\zps - \yps < 0.5$). On 2017 September 25 we obtained a 3.5 minutes follow-up $J$-band image with the FourStar camera \citep{persson2013} at the Magellan Baade telescope in Las Campanas Observatory.  This resulted in a flat $y-J=0.08\pm 0.13$ color, placing the candidate far from the brown dwarf color locus (see Figure 4 in \citealt{banados2016}) and making it a high priority for spectroscopic follow-up.

\begin{figure}
\includegraphics[width=\linewidth]{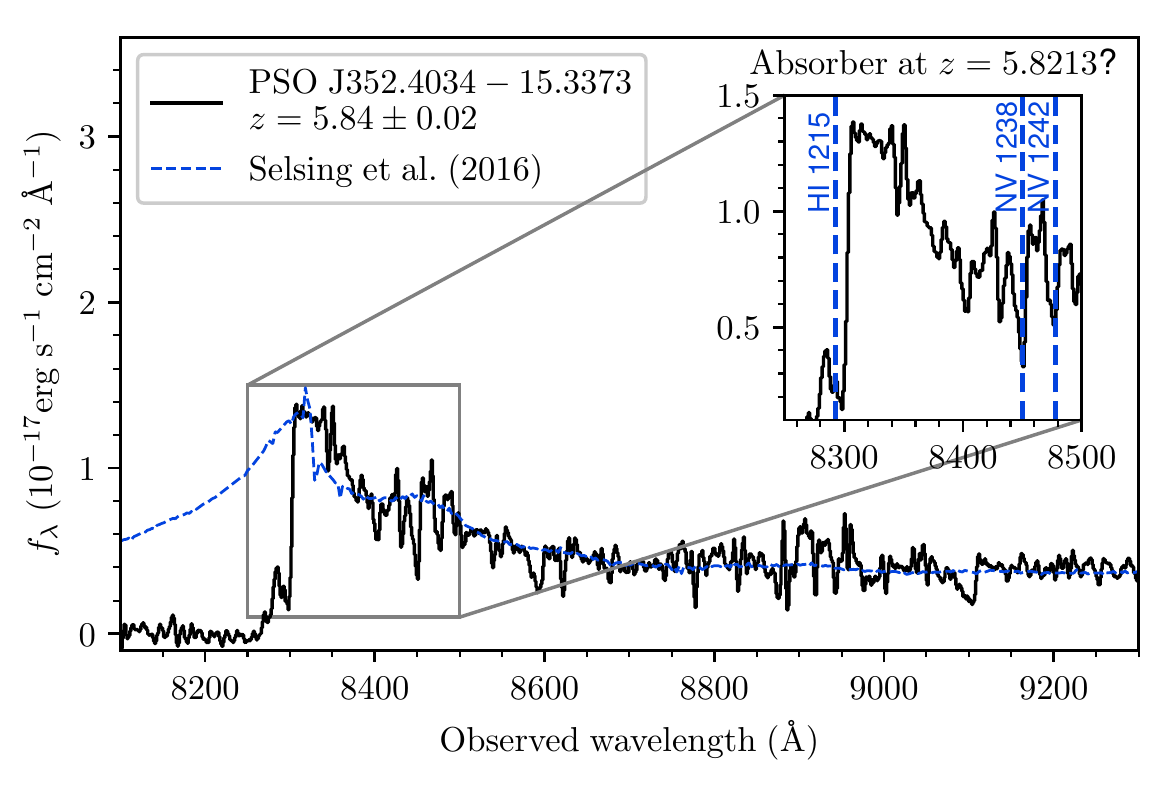}
\caption{Discovery spectrum of the radio-loud quasar \quasar\ at $z=5.84\pm 0.02$. The dashed line is the quasar template from \cite{selsing2016} at the redshift of the quasar for comparison. The inset shows a zoom-in indicating the tentative detection of an associated absorber at $z=5.8213$.    \label{fig:spectrum}}
\end{figure}

We confirmed \quasar\ as a quasar on 2017 September 26 with a 20 min spectrum taken with the Low-Dispersion Survey Spectrograph (LDSS3) at the Magellan Clay telescope in Las Campanas Observatory. The observations were carried out with the VPH-red grism and the  1\arcsec\ wide slit under good weather conditions and a seeing of $0\farcs7$. The data were reduced with standard routines including bias subtraction, flat fielding, sky subtraction, extraction, and wavelength calibration using He, Ne, and Ar lamps. The standard star Feige 110 \citep{moehler2014} was used for flux and telluric calibration. The final spectrum, scaled to match the \zps\ magnitude, is shown in Figure \ref{fig:spectrum}. The quasar's redshift was estimated by fitting the composite spectrum of bright $1<z<2$ quasars from \cite{selsing2016} to the data. The photometry and derived properties of the quasar \quasar\ are listed in Table \ref{tab:quasar}.

\subsection{Radio Surveys}

The quasar \quasar\ is coincident within 3\arcsec\ with a bright radio ($\sim$$15-100\,$mJy) source classified as a single unresolved source in the 1.4\,GHz NRAO VLA Sky Survey \citep[NVSS;][]{condon1998}, the GaLactic and Extragalactic All-sky MWA Survey \citep[GLEAM;][]{hurley-walker2017} using the Murchison Widefield Array \citep[MWA;][]{tingay2013} at $\sim$200\,MHz, and the TIFR GMRT Sky Survey \citep[TGSS;][]{intema2017} carried out at the Giant Metrewave Radio Telescope \citep[GMRT;][]{swarup1991} at $147.5\,$MHz.
However, at the large angular resolution of these surveys ($25\arcsec - 100\arcsec$) there are foreground galaxies in the field of view that could be contributing to the observed radio emission (see Figure~\ref{fig:radioimg}).

\subsection{Very Large Array (VLA) 3\,GHz Follow-up}

In order to confirm whether the radio emission is coming
from the quasar or other galaxies in the NVSS beam, we observed the
field with the Karl G. Jansky VLA on 2018 January
13 for one hour (36 minutes on-source).
We observed with the S-Band (2--4 GHz) using the 8-bit samplers and a correlator configuration that delivered 16 128\,MHz subbands, each with 64 channels (2 MHz/channel). We denote the central frequency of the S-band (3\,GHz) when referring to results obtained from these observations.
About 30\% of the band had to be flagged due to radio frequency
interference. Calibration was performed in CASA, using J2327--1447 as
the phase calibrator, and 3C48 to set the bandpass and absolute flux
density scale.

The calibrated data was CLEANed using Briggs weighting with a robust
factor of $-0.5$. The resulting resolution was $2\farcs6 \times 1\farcs4$.  The rms noise in the image is
20\,$\mu$Jy\,beam$^{-1}$.

\begin{figure}
\includegraphics[width=\linewidth]{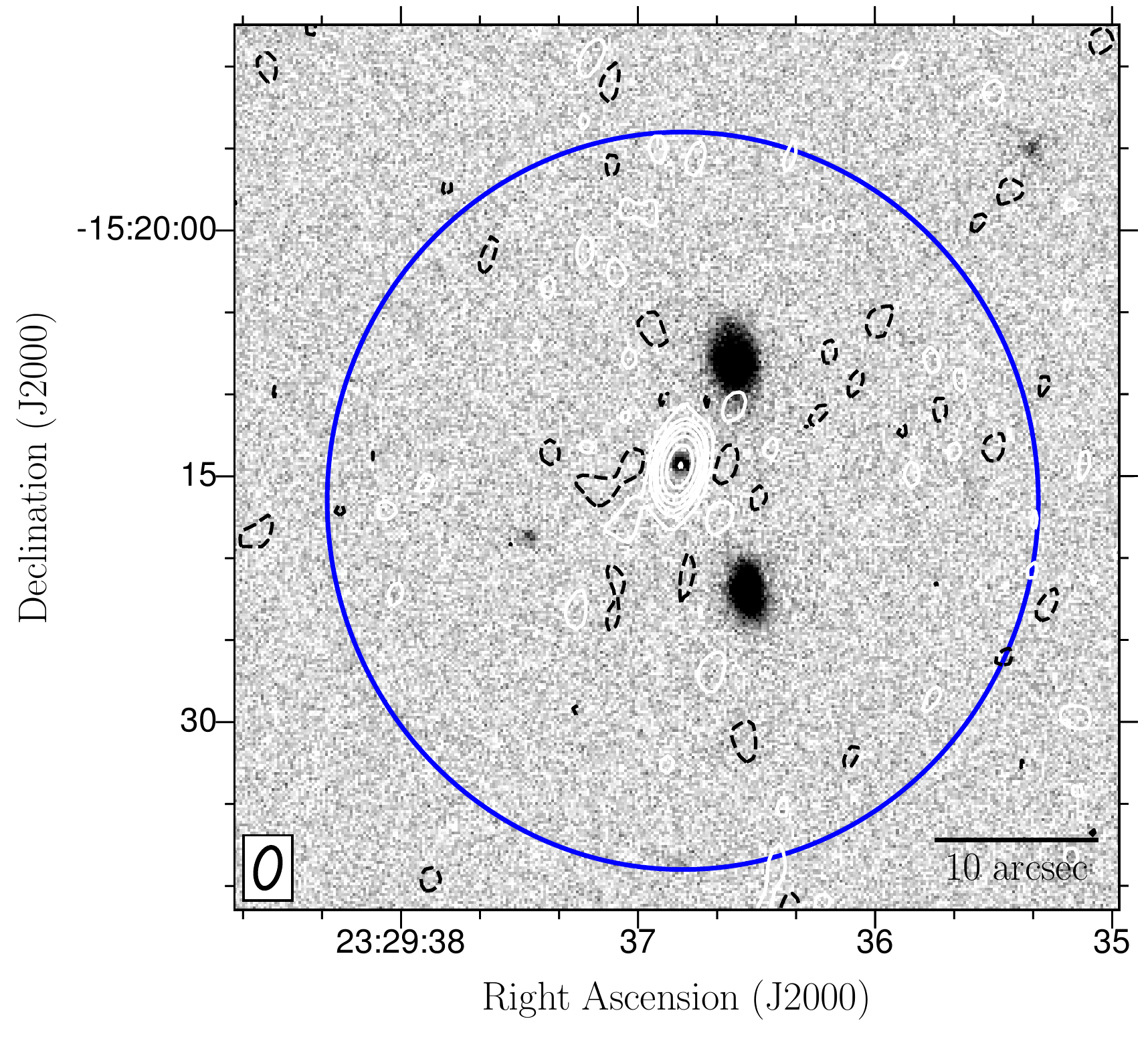}
\caption{Magellan/FourStar $J$-band image centered on the position of \quasar\ with north up and east left. The quasar is coincident with a $\sim$15\,mJy counterpart in the 1.4\,GHz NVSS radio survey. However, at the NVSS's resolution (as indicated by the blue circle) there are other galaxies in the field that could contribute to this emission. The contours show the follow-up VLA 3\,GHz observation of \quasar, confirming that all the radio emission is coming from the quasar. The rms noise in the image is
20\,$\mu$Jy$\,$beam$^{-1}$ and the contour levels are shown at ($-0.05$, $0.05$, $0.25$, $0.5$, $1.0$, $2.0$, $4.0$, $8.0$) mJy$\,$beam$^{-1}$. The VLA 3\,GHz $2\farcs6 \times 1\farcs4$ beam size is shown in the bottom-left panel. The source is unresolved with a peak flux density of 8.2\,mJy$\,$beam$^{-1}$, and a deconvolved size $\leq 0\farcs5$.  \label{fig:radioimg}}
\end{figure}

The observation resulted in a strong detection of the source consistent with the
near-infrared position of \quasar\ (see contours in Figure
\ref{fig:radioimg}) and no other radio source in the field.  Gaussian
fitting to the data implies that the source is
unresolved with a peak flux density of $8.20\pm 0.25\,$mJy
and a deconvolved size $\leq 0\farcs5$. The uncertainty in the flux density is dominated by the 3\% accuracy in the absolute flux calibration of 3C48 \citep{perley2013}.


\section{Results}

{The follow-up 3\,GHz VLA observation found no other radio sources within $50\arcsec$ from the quasar.
There is a $0.34 \pm 0.02$\,mJy radio source (the flux uncertainty does not include systematics on the flux calibration) at  R.A.=$23^{\rm h}29^{\rm m}33 \fs 999$, Decl.$=-15^{\circ}20^{\prime}43 \farcs 559$ ($50\farcs 3$ from the quasar). This faint radio source is well below the detection threshold of the shallow radio surveys listed in Section 2.2 and is not detected in our current optical or near-infrared images of the field.
}

\begin{figure}
\includegraphics[width=\linewidth]{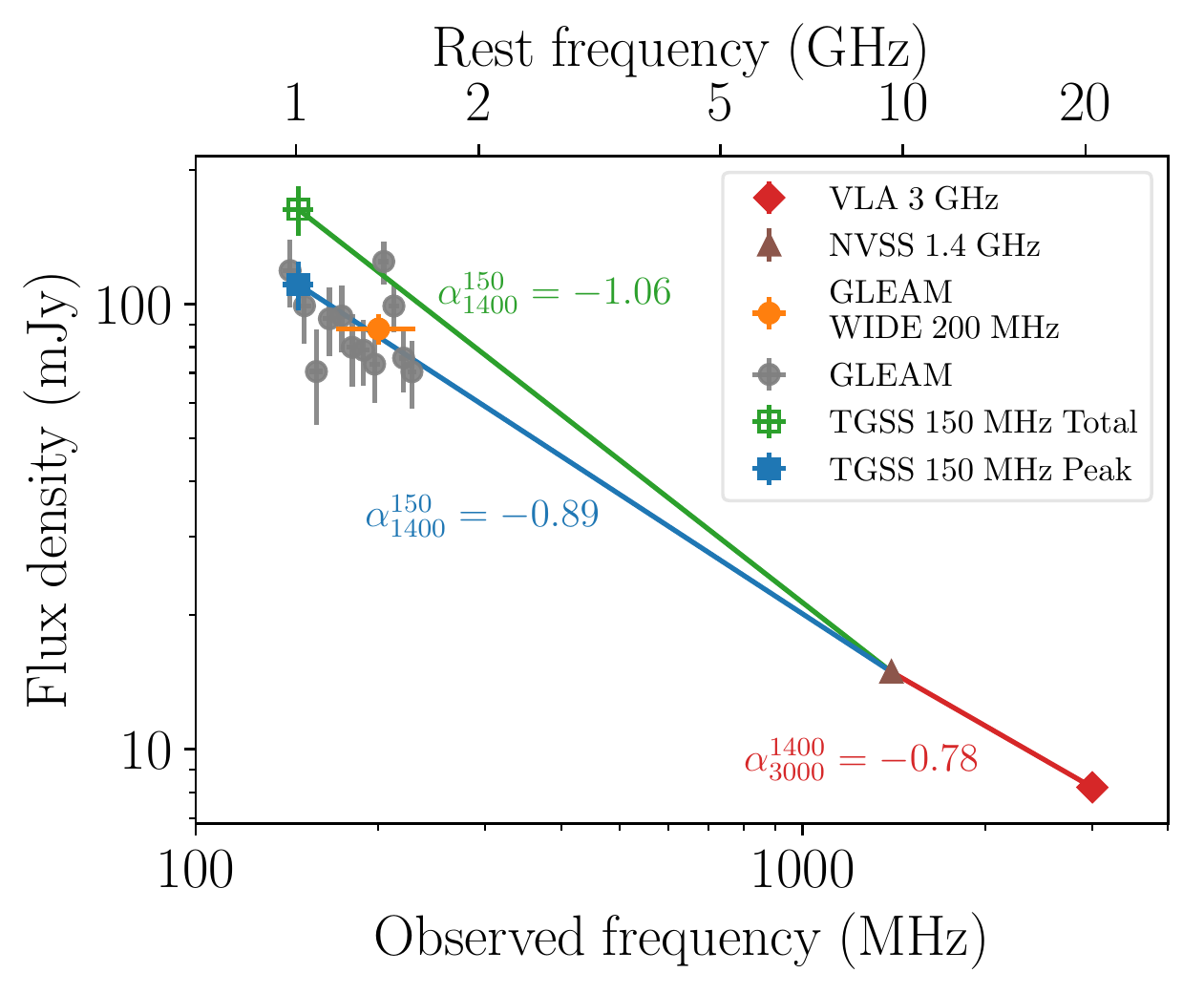}
\caption{Radio spectral energy distribution of \quasar, including data from our VLA 3\,GHz follow-up observation as well as from the NVSS, GLEAM, and TGSS surveys. We only show GLEAM detections with a signal-to-noise ratio greater than five.  The lower frequency measurements have a large scatter but they are consistent with a steep radio slope of $ -1.06 \lesssim \alpha_{1400}^{150}\lesssim -0.89$. The error bars for the 1.4 and 3\,GHz measurements are smaller than the plotted symbols.
\label{fig:sed}}
\end{figure}

Therefore, we can safely assume that the radio emission reported in the shallow public radio surveys is coming entirely from \quasar.  We show the radio spectral energy distribution of \quasar\ in Figure~\ref{fig:sed}.

The radio slope between the two VLA measurements at 3 and 1.4\,GHz is $\alpha^{1400}_{3000}= -0.78\pm0.07$. The GLEAM measurements show a large scatter but they are consistent with values between 80 and 120\,mJy at 200\,MHz \citep{hurley-walker2017}. Using the flux density of the GLEAM wide-band image, the radio slope between 1.4\,GHz and 200\,MHz is $\alpha_{1400}^{200}= -0.91\pm0.05$. The 150\,MHz TGSS total flux density reported in the catalog of \citet[][$163.1\pm 20.7\,$mJy]{intema2017}  is significantly different from the re-processed value of \citet[][$262\pm24\,$mJy]{degasperin2018}. Our own Gaussian fitting to the TGSS data is consistent with the values of \cite{intema2017}. The radio slope between 1.4\,GHz and 150\,MHz  is $\alpha_{1400}^{150}= -1.06\pm0.06$ if we consider the TGSS total flux density or $\alpha_{1400}^{150}= -0.89\pm0.06$ if we  consider the TGSS peak flux density.  We note that there are clear stripe-like artifacts in the TGSS image that could explain the discrepancies. It will therefore be important to obtain follow-up observations to robustly confirm the flux density at 150\,MHz.

\begin{figure}
\includegraphics[width=\linewidth]{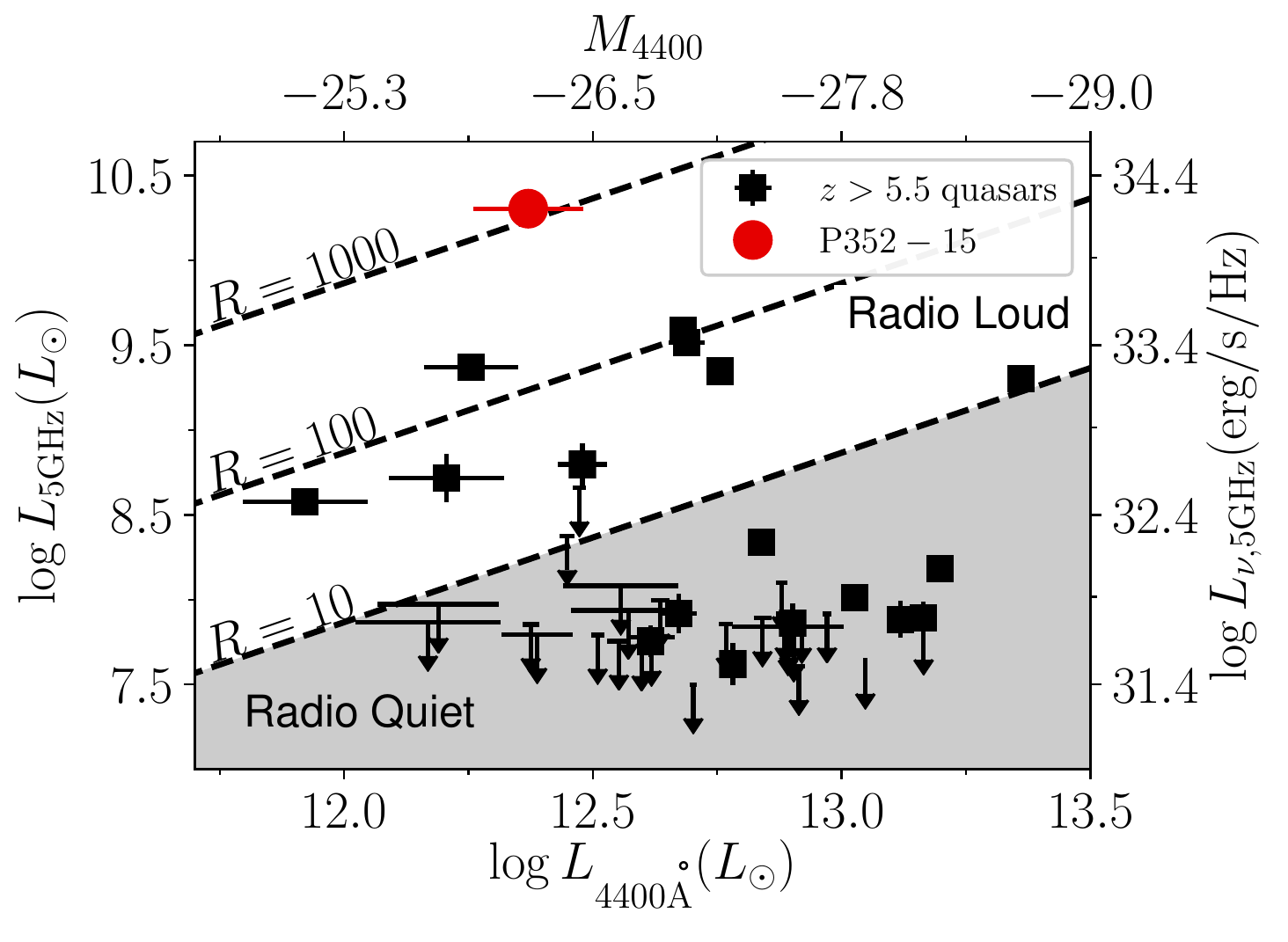}
\caption{Rest-frame 5\,GHz radio luminosity vs. rest-frame $4400\,$\AA\ optical luminosity for all redshift $z>5.5$ quasars detected at 1.4\,GHz or with strong limits (figure adapted from \citealt{banados2015a}). The arrows represent $3\sigma$ upper limits. The red circle corresponds to \quasar, the quasar discussed in this Letter. We show the average of the radio luminosity using two radio slopes ($\alpha=-0.89$ and $\alpha=-1.06$). We note that the symbol size is bigger in the y-axis than the whole range allowed by the two radio slopes (see also Table~1).
\label{fig:lums}}
\end{figure}

The radio loudness of an object is typically quantified as the radio/optical ratio between the rest-frame flux densities at 5\,GHz and $4400\,$\AA: $R=f_{\nu,5{\rm GHz}}/f_{\nu,4400{\rm A}}$. We estimate the flux density at $4400\,$\AA\ extrapolating the $1450\,$\AA\ flux density assuming an optical spectral index $\alpha_\nu=-0.5$ as in \cite{banados2015a}. The exact value of the radio loudness for \quasar\ depends on the radio spectral index, but using both our extreme cases ($\alpha=-1.06$ and $\alpha=-0.89$), we find $R\gtrsim 1000$, making it about one order of magnitude more radio loud than any other quasar known at these redshifts (Figure \ref{fig:lums}). We summarize the radio and optical properties of \quasar\ in Table 1.

Another point worth mentioning is that the quasar spectrum shows a possible associated absorption system in Ly$\alpha$ and \nv\  at $z=5.8213$ (Figure~\ref{fig:spectrum}).
If the absorption is confirmed with future observations, it may indicate a dense local environment.  Associated absorption could also be a probe of the interstellar medium of the host galaxy, often indicating strong outflows driven by radio-mode feedback associated with the expanding radio source \citep[e.g.,][]{vayner2017}.

\section{Concluding remarks}

The existence of this bright radio quasar at $z\sim 6$
is promising for future 21\,cm forest studies in the epoch of reionization \citep[e.g.,][]{semelin2016}. If the associated absorber is confirmed (Figure \ref{fig:spectrum}), \quasar\  could enable one of the key science cases of the Square Kilometre Array, \hi\ 21\,cm absorption spectroscopy in the epoch of reionization \citep{kanekar2004,morganti2015}, but in a much nearer future.
At $z\sim 5.8$, the 21\,cm line is shifted to $\sim 208\,$MHz, a frequency that can be studied with existing radio interferometers such as the MWA and the GMRT.

This radio-loud quasar will likely be an unprecedented laboratory for a number of studies, including the following. 

\begin{itemize}
\item  VLBA \citep{momjian2018} and X-ray studies of jet properties at the earliest cosmic times will provide a direct observational test for the effects of the cosmic microwave background on the most distant powerful jets  (\citealt{ghisellini2015}).

\item The role of radio-jets for the formation and growth of supermassive black holes \citep{volonteri2015}.

\item Search for direct evidence of radio-mode feedback as an important process during the formation of the first massive galaxies in the universe (\citealt{fabian2012}).

\end{itemize}

\floattable
\begin{table}[tbh]
\scriptsize
\centering
\caption{Properties and Photometry of the Radio-loud Quasar \quasar
}
\label{tab:quasar}
\begin{tabular}{lR}
\hline \hline
R.A. (J2000) & 23^{\rm h} 29^{\rm m} 36\fs8363   \\
Decl. (J2000) & -15^{\circ} 20^{\prime} 14 \farcs 460\\
Redshift  & 5.84 \pm 0.02 \\
$m_{1450,\mathrm{AB}}$   & 21.05 \pm 0.13  \\
$M_{1450}$   & -25.59 \pm 0.13  \\
$L_{\nu,1.4\,\mathrm{GHz}}$ ($\alpha=-1.06$) & (6.3 \pm 0.3) \times  10^{27} \mathrm{W}\,\mathrm{Hz}^{-1} \\
$L_{\nu,1.4\,\mathrm{GHz}}$ ($\alpha=-0.89$) & (4.5 \pm 0.2) \times  10^{27} \mathrm{W}\,\mathrm{Hz}^{-1} \\
$L_{5\,\mathrm{GHz}}$ ($\alpha=-1.06$) &10^{10.33 \pm 0.02} L_\odot \\
$L_{5\,\mathrm{GHz}}$ ($\alpha=-0.89$) &10^{10.28 \pm 0.02} L_\odot \\
$R$ ($\alpha=-1.06$) & 1234 \pm 317 \\
$R$ ($\alpha=-0.89$) & 1105 \pm 284 \\
$L_{4400\,\mathrm{A}}$ & 10^{12.37 \pm 0.11} L_\odot \\
\hline
Pan-STARRS1 $\ips$ & 1.8 \pm 0.5 \,\mu \mathrm{Jy} \\
\ensuremath{i_{\rm P1, AB}} & 23.28 \pm 0.28 \\
Pan-STARRS1 $\zps$ & 11.8 \pm 0.7 \,\mu \mathrm{Jy} \\
\ensuremath{z_{\rm P1, AB}} & 21.22 \pm 0.07 \\
Pan-STARRS1 $\yps$ & 13.6 \pm 1.5 \,\mu \mathrm{Jy} \\
\ensuremath{y_{\rm P1, AB}} & 21.07 \pm 0.13 \\
FourStar $J$  & 14.6 \pm 0.4 \,\mu \mathrm{Jy} \\
$J_{\rm AB}$  & 20.99 \pm 0.03  \\
VLA-S $3.0\,$GHz  & 8.20 \pm 0.25 \, \mathrm{mJy} \\
NVSS $1.4\,$GHz  & 14.9 \pm 0.7 \, \mathrm{mJy} \\
GLEAM WIDE $200\,$MHz  & 87.8 \pm 6.9 \, \mathrm{mJy} \\
TGSS $150\,$MHz Peak  & 110.6 \pm 13.8 \, \mathrm{mJy} \\
TGSS $150\,$MHz Total  & 163.1 \pm 20.7 \, \mathrm{mJy} \\
\hline
\end{tabular}
\end{table}

\acknowledgments
We thank the referee for suggestions that improved the clarity of this Letter.
We thank Daniel D.~Kelson for his support with the FourStar data.
F.W. and B.P.V. acknowledge funding through ERC grants ``Cosmic Dawn'' and ``Cosmic Gas.'' 
This Letter includes data gathered with the 6.5 m Magellan Telescopes located at Las Campanas Observatory, Chile.
We are grateful to the VLA staff for providing DDT observations for this program.
The National Radio Astronomy Observatory is a facility of the National Science Foundation operated under cooperative agreement by Associated Universities, Inc.
The Pan-STARRS1 Surveys (PS1) and the PS1 public science archive have been made possible through contributions by the Institute for Astronomy, the University of Hawaii, the Pan-STARRS Project Office, the Max-Planck Society and its participating institutes, the Max Planck Institute for Astronomy, Heidelberg and the Max Planck Institute for Extraterrestrial Physics, Garching, The Johns Hopkins University, Durham University, the University of Edinburgh, the Queen's University Belfast, the Harvard-Smithsonian Center for Astrophysics, the Las Cumbres Observatory Global Telescope Network Incorporated, the National Central University of Taiwan, the Space Telescope Science Institute, the National Aeronautics and Space Administration under grant No. NNX08AR22G issued through the Planetary Science Division of the NASA Science Mission Directorate, the National Science Foundation grant No. AST-1238877, the University of Maryland, Eotvos Lorand University (ELTE), the Los Alamos National Laboratory, and the Gordon and Betty Moore Foundation.

%


\facilities{PS1 (GPC1), Magellan:Clay (LDSS3 spectrograph), Magellan:Baade (FourStar), VLA}


\software{astropy \citep{astropy2018}, matplotlib \citep{hunter2007},
CASA \citep{mcmullin2007}}

\end{document}